\documentclass[11pt, a4paper]{article}
\pdfoutput=1

\usepackage{jcappub}

\usepackage{comment}
\usepackage{amsmath}
\usepackage{amsfonts}
\usepackage{amssymb}
\usepackage{graphicx}
\usepackage{mathrsfs}
\usepackage{latexsym}
\usepackage{color}
\usepackage{slashed, cancel}
\usepackage{hyperref}
\usepackage{cleveref}
\usepackage{float}
\usepackage{tikz}
\usepackage{soul} 
\usepackage{mathtools}
\usepackage[utf8]{inputenc} 
\usepackage{csquotes}
\usepackage{subcaption}
\usepackage{enumitem}
\usepackage{journals}
\allowdisplaybreaks
\usepackage{multirow}
\usepackage{pifont}
\usepackage{orcidlink}

\hypersetup{urlcolor=blue, colorlinks=true} 

\usepackage{fancyhdr}
\numberwithin{equation}{section}

\definecolor{orange}{rgb}{1,0.4,0}
\definecolor{green}{rgb}{0,0.65,0}
\definecolor{rossos}{rgb}{0.8,0.2,0.3}
\definecolor{bluscuro}{rgb}{0.15, 0.2, .85}
\definecolor{bluchiaro}{cmyk}{1,.3,0.,0.1}
\hypersetup{colorlinks, citecolor=bluscuro, linkcolor=bluscuro, urlcolor=bluscuro}

\makeatother   
\newcommand{\GeV}{{\rm \,GeV}}

\newcommand{\MeV}{{\rm \,MeV}}

\newcommand{\Msun}{M_\odot}

\newcommand{\Rstar}{R_\star}
\newcommand{\vstar}{v_\star}

\newcommand{\fFD}{f_{\rm FD}}

\newcommand{\mbeff}{m_i^{\rm eff}}

 \def\be   {\begin{equation}}   \def\ee   {\end{equation}}
 \def\ba   {\begin{array}}      \def\ea   {\end{array}}
 \def\bea  {\begin{eqnarray}}   \def\eea  {\end{eqnarray}}
 \def\bean {\begin{eqnarray*}}  \def\eean {\end{eqnarray*}}
 \def\nn{\nonumber}

\begin{document}

\hfill ADP-21-11/T1158


\title{Using neutron stars to probe dark matter charged under a $L_\mu-L_\tau$ symmetry}

\author[a]{Nicole F.\ Bell\,\orcidlink{0000-0002-5805-9828},}
\author[b]{Giorgio Busoni\,\orcidlink{0000-0002-8527-0768},}
\author[a]{Avirup Ghosh\,\orcidlink{0000-0002-4781-842X}}

\affiliation[a]{ARC Centre of Excellence for Dark Matter Particle Physics, \\
School of Physics, The University of Melbourne, Victoria 3010, Australia}
\affiliation[b]{ARC Centre of Excellence for Dark Matter Particle Physics, \\
Department of Physics, University of Adelaide, Adelaide, SA 5005, Australia}

\emailAdd{n.bell@unimelb.edu.au}
\emailAdd{giorgio.busoni@adelaide.edu.au}
\emailAdd{avirup.ghosh@unimelb.edu.au}

\abstract{Kinetic heating of old cold neutron stars, via the scattering of dark matter with matter in the star, provides a promising way to probe the nature of dark matter interactions. We consider a dark matter candidate that is a Standard Model singlet Dirac fermion, charged under a $U(1)_{L_\mu-L_\tau}$ symmetry. Such dark matter interacts with quarks and electrons only via loop-induced couplings, and hence is weakly constrained by direct-detection experiments and cosmic-microwave background observations. However, tree-level interactions with muons enable the dark matter to interact efficiently with the relativistic muon component of a neutron star, heating the star substantially. Using a fully relativistic approach for dark matter capture in the star, we show that observations of old cold neutron stars can probe a substantial, yet unexplored, region of parameter space for dark matter masses in the range 100 MeV - 100 GeV.}

\maketitle

\section{Introduction}
\label{sec:intro}

Understanding the nature of particle dark matter (DM) is one of the most pressing issues for the particle physics community today. While direct detection efforts that search for evidence of DM-induced nuclear- or electronic- recoil signals is the gold standard approach, this technique has inherent shortcomings for particular types of DM. Specifically, signals in these experiments are suppressed for many well-motivated types of DM interactions, including those that give rise to spin- or momentum-dependent scattering, inelastic interactions, and some types of leptophilic interactions. In such cases, DM capture in stars~\cite{Gould:1987ju,Gould:1987ir,Jungman:1995df,Kumar:2012uh,Kappl:2011kz,Busoni:2013kaa,Bramante:2017xlb,Busoni:2017mhe}, and compact astrophysical objects such as White Dwarfs (WDs)~\cite{Bertone:2007ae,McCullough:2010ai,Hooper:2010es,Amaro-Seoane:2015uny,Panotopoulos:2020kuo,Curtin:2020tkm,Bell:2021fye,Bell:2024qmj} and Neutron Stars (NSs)~\cite{Goldman:1989nd,Bell:2018pkk,Bell:2019pyc} offer valuable astrophysical probes that avoid the suppressions relevant for terrestrial experiments.

While DM capture in stars can lead to various consequences, we shall focus on the DM-induced heating of old cold NSs~\cite{Baryakhtar:2017dbj,Raj:2017wrv,Bell:2018pkk,Camargo:2019wou,Bell:2019pyc,Garani:2019fpa,Acevedo:2019agu,Joglekar:2019vzy,Gonzalez,Maity:2021fxw}. 
When the DM capture rate in a NS is maximised, the energy transferred during the capture and subsequent thermalisation of the DM can heat a NS to temperatures of $\sim 2000$~K, which we may be able to probe with upcoming infrared telescopes~\cite{Baryakhtar:2017dbj}. Given the enormous mass and density of a typical NS, the probability of DM capture within the star can become of order unity for DM-neutron scattering cross-sections of ${\cal O}( 10^{-45}\,{\rm cm}^2)$~\cite{Bell:2020jou,Bell:2020obw,Anzuini:2021lnv}. This is comparable to the typical scale that is probed by DM direct-detection experiments, in the optimal case where the DM-nucleon interaction is spin-independent and the DM mass is in the GeV range.
Importantly, $\sigma \sim {\cal O}( 10^{-45}\,{\rm cm}^2)$ remains the scale at which NS capture is efficient even for momentum-suppressed interactions or leptophilic DM~\cite{Bell:2019pyc,Bell:2020lmm}. In the case of the latter, this offers a potential probe of DM-lepton interactions that is more sensitive than current electron-recoil direct detection experiments.    

In this paper, we consider a model where DM has tree-level interaction only with the second and third generation leptons. Specifically, we assume that the Standard Model (SM) is extended by a $U(1)_{L_\mu-L_\tau}$ gauge group, broken at some high-scale, under which the DM is charged. The $U(1)_{L_\mu-L_\tau}$ symmetry provides an attractive anomaly-free extension of the SM~\cite{Foot:1990mn,He:1991qd}, which is frequently invoked to explain the long-standing issues of the muon anomalous magnetic moment and non-zero neutrino mass~\cite{Foot:1990mn,He:1991qd,Foot:1994vd,Heeck:2011wj,Bell:2000vh}. In this model, the DM interacts with SM particles only via its coupling to the $U(1)_{L_\mu-L_\tau}$ gauge boson, $Z'$. This scenario would be challenging to probe with direct detection experiments, because the $Z'$ does not couple directly to nucleons or electrons. Instead, scattering in direct detection experiments would occur only via kinetic mixing of the $Z'$ with the SM photon. We shall demonstrate that the scattering of this $U(1)_{L_\mu-L_\tau}$-charged DM with the small relativistic muon component of a NS can provide an important probe of this model.

DM capture in NSs has been treated with varying degrees of sophistication. While early calculations were modelled on those for DM capture in the Sun, improved treatments that correctly account for the extreme physics of a NS environment have been developed in recent years~\cite{Bell:2020jou,Bell:2020lmm,Bell:2020obw,Anzuini:2021lnv}. This incorporates the composition of the NS, including full radial dependence of the densities and the chemical potentials, relativistic scattering targets, Pauli blocking,  gravitational focusing of the DM trajectories, and the opacity of the star. In the case of scattering on the hadronic targets, this treatment also incorporates hadronic form factors for quasi-relativistic scattering, and the strong self-interactions of the hadrons.

It is important to note that NSs contain appreciable electron and muon components, with large chemical potentials and relative abundances determined via beta-equilibrium in the interior of the star. A fully relativistic treatment is essential for DM scattering from these highly degenerate, relativistic, lepton species~\cite{Joglekar:2019vzy,Joglekar:2020liw,Bell:2020lmm}. The capture of leptophilic DM in NSs has previously been considered in a number of model independent studies that parameterised the interaction as an effective operator, assuming new physics at some high energy scale. The results of those studies cannot be directly applied to the case of a light mediator, such as that studied here. A NS analysis of $L_\mu-L_\tau$-charged DM was previously performed in Ref.~\cite{Garani:2019fpa}, for the case of scalar DM. However, this previous work did not perform a relativistic treatment of the scattering, as the importance of this feature for DM-lepton scattering had not yet been fully appreciated.

In this paper, we perform a full relativistic treatment of NS capture for a fermionic DM interacting via an $U(1)_{L_\mu-L_\tau}$ portal \cite{Baek:2008nz,Hapitas:2021ilr}. We shall compare our estimates for the NS sensitivity with existing constraints arising from collider experiments, CMB observations, DM direct detection experiments, and the requirement that the correct DM relic density be obtained. While the DM has tree-level interactions only with $\mu$, $\nu_\mu$, $\tau$ and $\nu_\tau$ via the exchange of the $Z'$ gauge boson, interactions with other SM particles will be generated through the $Z'$-photon kinetic mixing. Considering reasonable values of the kinetic mixing parameter (which we assume to be generated at the $U(1)_{L_\mu-L_\tau}$ breaking scale) we delineate the allowed region of the DM parameter space spanned by DM $L_\mu-L_\tau$ charge $Q_\chi$ and DM mass $m_\chi$, for two benchmark values of the $U(1)_{L_\mu-L_\tau}$ gauge coupling and $Z^\prime$ mass. Our results show that NS capture can potentially test unexplored regions of the $L_\mu-L_\tau$-portal DM parameter space, including the phenomenologically interesting region where $Q_\chi \sim 1$, for all DM masses in the range 100 MeV - 100 GeV.

This paper is organised as follows: In Section~\ref{sec:mutauModel}, we discuss the $U(1)_{L_\mu-L_\tau}$-charged DM model and present existing constraints on the coupling of the $Z'$ to SM particles and to DM. In Section~\ref{sec:ns} we outline the DM capture rate calculations and determine the NS sensitivity to $U(1)_{L_\mu-L_\tau}$ charged DM. We summarise and conclude in Section~\ref{sec:conclusion}.


\section{$L_\mu-L_\tau$ portal dark matter model and existing constraints}
\label{sec:mutauModel}

We assume that the SM is extended by the addition of a $U(1)_{L_\mu-L_\tau}$ gauge group, and a Dirac fermion dark matter candidate, $\chi$.  We assign the $\{\mu,\nu_\mu\}$, $\{\tau, \nu_\tau\}$, and $\chi$ fields $L_\mu-L_\tau$ charges of 1, -1, and $Q_\chi$ respectively. The relevant low-energy Lagrangian is then given by~\cite{Baek:2008nz,Hapitas:2021ilr}
\begin{eqnarray}
\mathcal{L} &=& \mathcal{L}_{\rm SM} + \mathcal{L}_{\rm \mu-\tau} + \mathcal{L}_{\rm \chi}\\
\mathcal{L}_{\rm \mu-\tau} &=&  - \frac{1}{4}F_{\mu\tau}^{\alpha\beta}F_{\mu\tau\,\alpha\beta} + \frac{\varepsilon_0}{2}F_{\mu\tau}^{\alpha\beta}B_{Y\alpha\beta} + \frac{1}{2}m^2_{Z^\prime}Z^{\prime\,\alpha}Z^{\prime\,\beta} +  J_{Z^\prime}^\alpha Z^\prime_\alpha\\
J_{Z^\prime}^\alpha &=& g_{\mu\tau}\left(\bar{\mu}\gamma^\alpha \mu - \bar{\tau}\gamma^\alpha \tau + \bar{\nu}_\mu \gamma^\alpha \nu_\mu - \bar{\nu}_\tau \gamma^\alpha \nu_\tau \right),\\
\label{eq:LagLmuLtaumain}
\mathcal{L}_\chi &=& \bar{\chi}\left(i\cancel{\partial}- m_\chi\right)\chi + Q_\chi g_{\mu\tau} \bar{\chi}\gamma^\alpha \chi Z^\prime_\alpha,
\label{eq:LagLmuLtauDM}
\end{eqnarray}
where $g_{\mu\tau}$ is the $U(1)_{L_\mu-L_\tau}$ gauge coupling, $Z^\prime$ is the 
$L_\mu-L_\tau$ gauge boson, $F_{\mu\tau}^{\alpha\beta} = \partial^\alpha Z^{\prime\,\beta} - \partial^\beta Z^{\prime\,\alpha}$ is the $U(1)_{L_\mu-L_\tau}$ field strength tensor and $B_{Y\alpha\beta}$ is the field strength tensor for the SM hypercharge gauge group $U(1)_Y$. 
We assume that the $U(1)_{L_\mu-L_\tau}$ symmetry has undergone spontaneous symmetry breaking at some high scale $\Lambda_{\mu\tau} \sim \mathcal{O}({\rm TeV})$, and hence the $Z^\prime$ gauge boson is always massive for the purpose of this study. 

After the $U(1)_{L_\mu-L_\tau}$ is broken, there may be kinetic mixing of the $Z^\prime$ with the other neutral gauge bosons. Under the assumption that all BSM particles charged under both the $U(1)_{L_\mu-L_\tau}$ and the SM $U(1)_Y$ are much heavier than $\Lambda_{\mu\tau}$, no appreciable kinetic mixing with hypercharge is generated and thus $\varepsilon_0=0$. However, while its value at energies above the EW scale is encoded as $\varepsilon_0$, the kinetic mixing is momentum dependent. It will be induced at low energy at the one-loop level, because $\mu,\tau,\nu_{\mu,\tau}$ are charged under both $U(1)_{L_\mu-L_\tau}$ and $U(1)_Y$. The loop-induced kinetic mixing is given by~\cite{Hapitas:2021ilr}
\begin{equation}
\varepsilon(Q^2) = \varepsilon_0-\frac{e g_{\mu\tau}}{2\pi^2} \int_0^1 dx x(1-x) \log \left[\frac{m^2_\tau+x(1-x)Q^2}{m^2_\mu+x(1-x)Q^2} \right],
\label{eq:kinmixoneloop}
\end{equation}
where $Q^2$ represents the invariant four-momentum transfer involved in the 
$Z^\prime-Y$ two-point correlation function. It is easy to verify that this induced 
kinetic mixing shows flat asymptotic behaviour both at $Q^2\gg m_\tau^2$ and $Q^2\ll m_\mu^2$, 
given by,
\begin{eqnarray}
    \varepsilon(Q^2\gg m_\tau^2) &=&  \varepsilon_0,\\
    \varepsilon(Q^2\ll m_\mu^2) &=& \varepsilon_{IR} = \varepsilon_0-\frac{e g_{\mu\tau}}{12\pi^2} \log \left[\frac{m^2_\tau}{m^2_\mu} \right].
\end{eqnarray}
After performing a field transformation to diagonalise the gauge field 
kinetic terms, the $Z^\prime$ obtains dark-photon-like couplings to the 
SM fermions, and the $J_{Z^\prime}^\alpha$ is modified to
\begin{eqnarray}
J_{Z^\prime}^\alpha \rightarrow J_{Z^\prime}^\alpha + \varepsilon J_{em}^\alpha 
\end{eqnarray}
where $J_{em}^\alpha = e\,Q_f\,\bar{f}\gamma^\alpha f$, and $f$ is a SM fermion 
carrying electromagnetic charge $Q_f$. To avoid strong existing constraints on dark photons, we will assume that $e\varepsilon_0\sim \mathcal{O} (\alpha g_{\mu\tau})$. This ensures that dark-photon-like interactions induced by kinetic mixing are suppressed by $\alpha$ compared to tree-level $U(1)_{\mu-\tau}$ specific interactions. A specific choice that satisfies this requirement and that has a significant impact on phenomenology is $\varepsilon_{IR}=0$, because for this choice all couplings to SM fermions, except $\mu,\tau,\nu_\mu,\nu_\tau$, vanish at zero momentum transfer. Therefore, we will consider two possible choices for the kinetic mixing parameter: $\varepsilon_0=0$ and $\varepsilon_{IR}=0$ (which corresponds to $\varepsilon_0=\frac{e g_{\mu\tau}}{12\pi^2} \log \left[\frac{m^2_\tau}{m^2_\mu} \right]$).

\subsection{Existing constraints on the $L_\mu-L_\tau$ gauge boson}
\label{sec:SMcons}

\begin{figure}[tb]
\centering
\includegraphics[width=9.35cm,height=7.5cm]{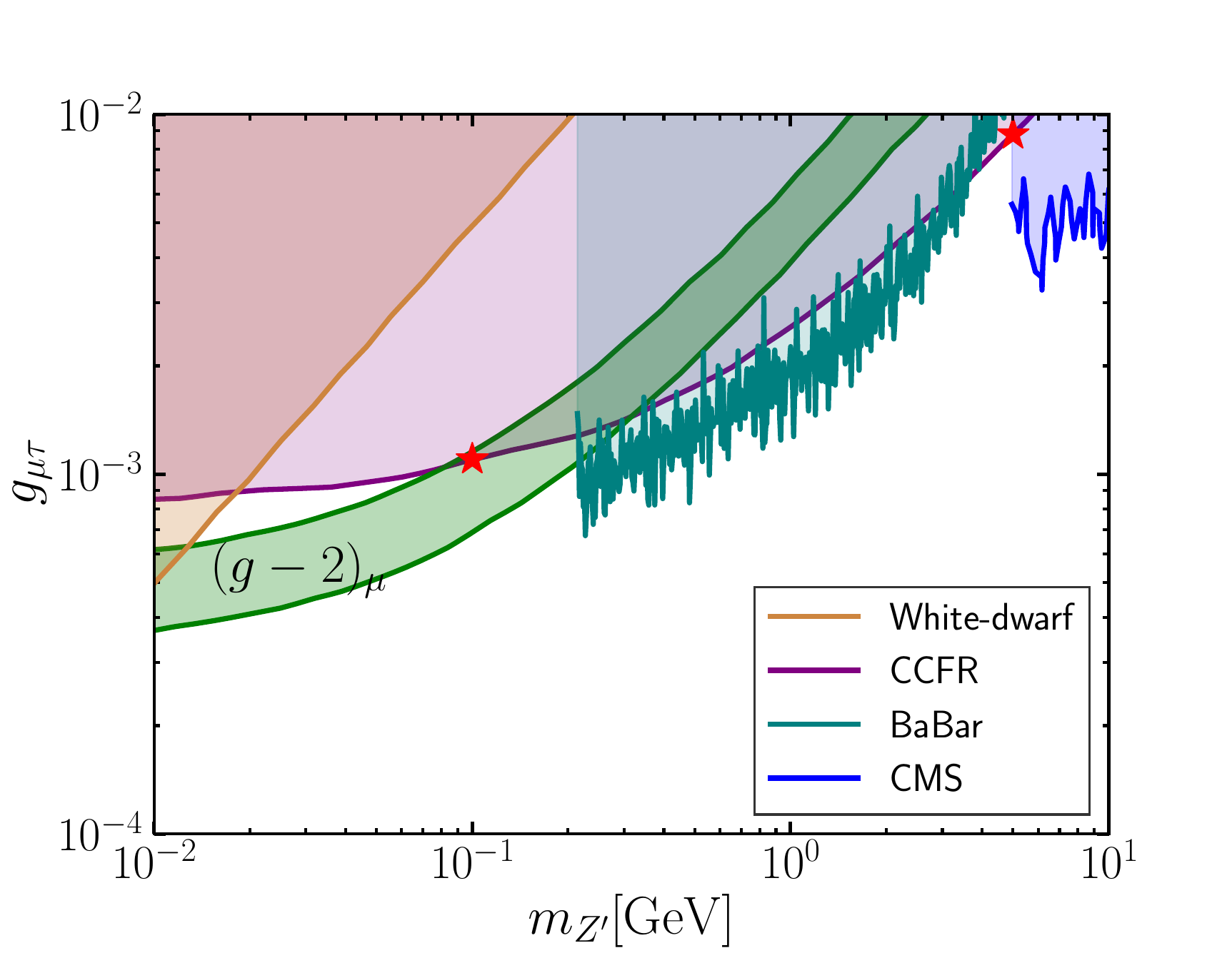}
\caption{Status of existing constraints on the $L_\mu-L_\tau$ model, in the $g_{\mu\tau}-m_{Z^\prime}$. The CCFR~ \cite{PhysRevLett.66.3117,Altmannshofer:2014pba}, BaBar~\cite{BaBar:2016sci} and CMS~\cite{CMS:2018yxg} constraints do not depend on the choice of $\varepsilon_0$ and are shown as the purple, dark green and blue shaded regions, respectively. The region favoured by $(g-2)_\mu$ measurements~\cite{Keshavarzi:2021eqa} also does not depend on $\varepsilon_0$ and is shown in light green. The region excluded by the observation of White Dwarf cooling~\cite{Dreiner:2013tja} depends on $\varepsilon_0$ and is shown as shaded brown for $\varepsilon_0=0$, which is the most constraining case in the $m_{Z^\prime}$ range that is considered. The red stars represent two benchmark points we shall use in our NS analysis.}
\label{fig:Zprimecons}
\end{figure}

Given the dominant interactions of the $Z^\prime$s with the SM muons and taus, several experiments are quite sensitive to $L_\mu-L_\tau$ gauge coupling $g_{\mu\tau}$. First, let us discuss some of these experimental constraints which are relevant for the parameter space under consideration. Note that, under the approximation $e\varepsilon_0\sim \mathcal{O} (\alpha g_{\mu\tau})$, these constraints receive sub-leading contributions from the kinetic mixing and thus depend dominantly on $g_{\mu\tau}$:

\begin{itemize}
\item \textbf{$\nu$-trident production:} In presence of $Z^\prime$-mediated interactions between muons and neutrinos, 
the rate of $\nu$-trident production, i.e., 
$\nu N \rightarrow \nu N \mu^+ \mu^-$ is expected to be enhanced. 
Using a neutrino beam of energy $\sim 160\,{\rm GeV}$ and an iron 
target, the Columbia-Chicago-Fermilab-Rochester (CCFR) neutrino 
experiment at Fermilab has set stringent limits on the rate of such 
processes~\cite{PhysRevLett.66.3117}. Utilising these results, 
Ref.~\cite{Altmannshofer:2014pba} has constrained the $g_{\mu\tau}-m_{Z^\prime}$ parameter space, as indicated by the purple-shaded region in Fig.~\ref{fig:Zprimecons}.  

\item \textbf{BaBar:} The BaBar collaboration 
has conducted a search for $e^+e^- \rightarrow \mu^+\mu^- \mu^+\mu^-$, using $e^+-e^-$ collisions at $\sim 10.7\,{\rm GeV}$ center-of-mass energy and $514\,{\rm fb}^{-1}$ of luminosity, to derive an upper-limit on $g_{\mu\tau}$ for $Z^\prime$ masses in the range 0.2 - 10 GeV~\cite{BaBar:2016sci}. This constraint is indicated by the dark-green shaded region in Fig.~\ref{fig:Zprimecons}. 

\item \textbf{CMS:}  The CMS collaboration has performed a search for $4\mu$ events, using the data from $pp$ collisions at 13~TeV center-of-mass energy and 77.3${\rm fb}^{-1}$ integrated luminosity. This sets an upper-limit on $g_{\mu\tau}-m_{Z^\prime}$ for  $m_{Z^\prime} \gtrsim 5\,{\rm GeV}$~\cite{CMS:2018yxg}, as shown in blue in Fig.~\ref{fig:Zprimecons}. 

\item \textbf{Muon $g-2$:} 
Due to the presence of the $\mu\bar{\mu}Z^\prime$ coupling, one-loop diagrams involving an off-shell $\mu$ and $Z^\prime$ can give rise to an additional contribution to the Muon anomalous magnetic moment, 
$a_\mu$. There is a longstanding discrepancy between the SM prediction, $a^{\rm SM}_\mu$ ~\cite{Aliberti:2025beg}, and the experimentally measured value $a^{\rm expt}_\mu$. The additional contribution to $a_\mu$, induced by the $Z'$ coupling, is given  by~\cite{Hapitas:2021ilr,Baek:2001kca}:
\begin{eqnarray}
\Delta a_\mu &=& \frac{(g_{\mu\tau}+e\varepsilon)^2}{4\pi^2} \int_0^1 dx \frac{m^2_\mu\,x^2(1-x)}{m^2_{Z^\prime}(1-x)+m^2_\mu\,x^2}.
\label{eq:muongm2}
\end{eqnarray}
The parameter space consistent with the latest measurements of  the muon anomalous magnetic moment $(g-2)_\mu$~\cite{Keshavarzi:2021eqa} is shown by the green band in Fig.~\ref{fig:Zprimecons}. 
\end{itemize}

In Figure~\ref{fig:Zprimecons}, we have also marked two benchmark points that we adopt for our NS analysis below. In addition, Fig.~\ref{fig:Zprimecons} also includes a constraint arising from the observations of white-dwarf cooling (shown by brown-shaded region) due to the decay of plasmons to neutrinos via $Z^\prime$-mediated processes~\cite{Dreiner:2013tja}. Note that this constraint depends on the value of $\varepsilon_0$ -- we have presented the limit obtained for $\varepsilon_0 = 0$; for $\varepsilon_{\rm IR}=0$, the limit is slightly relaxed (see Ref.~\cite{Hapitas:2021ilr}).

\subsection{Dark Matter related constraints} 
\label{sec:DM}

The constraints on the $L_\mu-L_\tau$ model discussed above are all independent of a possible coupling to DM. We now discuss the additional requirements that must be met if we introduce the coupling of the $Z'$ to the DM field, $\chi$, as described by Eq.~\ref{eq:LagLmuLtauDM}.  Specifically, DM direct searches and cosmic microwave background (CMB) observations place strong constraints on the DM parameter space, which we outline below.  We shall also determine the mass and coupling parameters for which the correct DM relic density is obtained through standard thermal freeze-out. Note that the relic density and CMB constraints are dominated by processes involving only $\chi$, $\mu$, $\tau$ and $\nu_{\mu,\tau}$, and are therefore independent of the choice of $\varepsilon_0$, while direct-detection constraints depend on the coupling to quarks and are therefore sensitive to $\varepsilon_0$.

\begin{figure}[htb!]
\centering
\includegraphics[width=8.4cm,height=6.5cm]{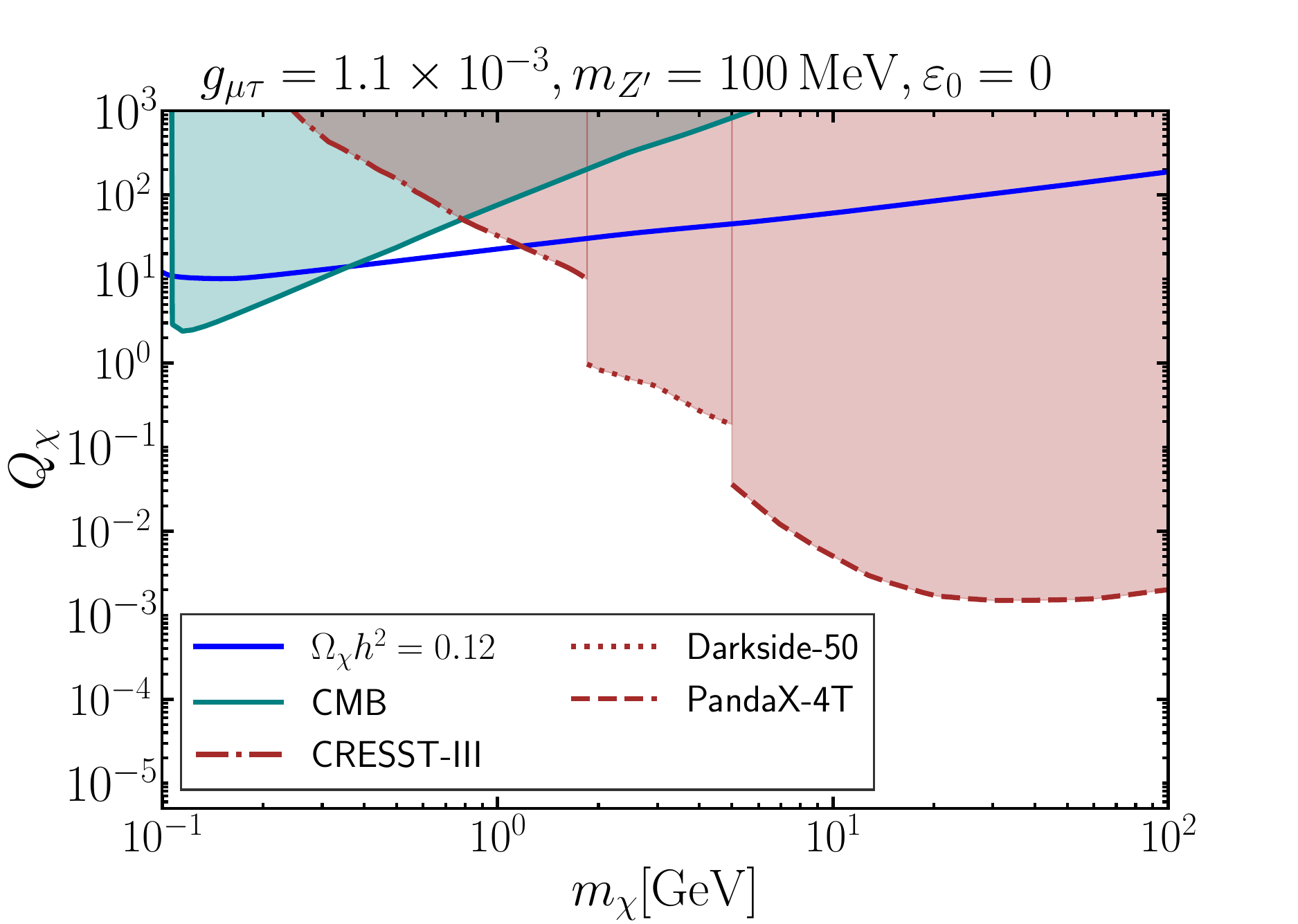}\!\!\!\!\!\!\!\!\!
\includegraphics[width=8.4cm,height=6.5cm]{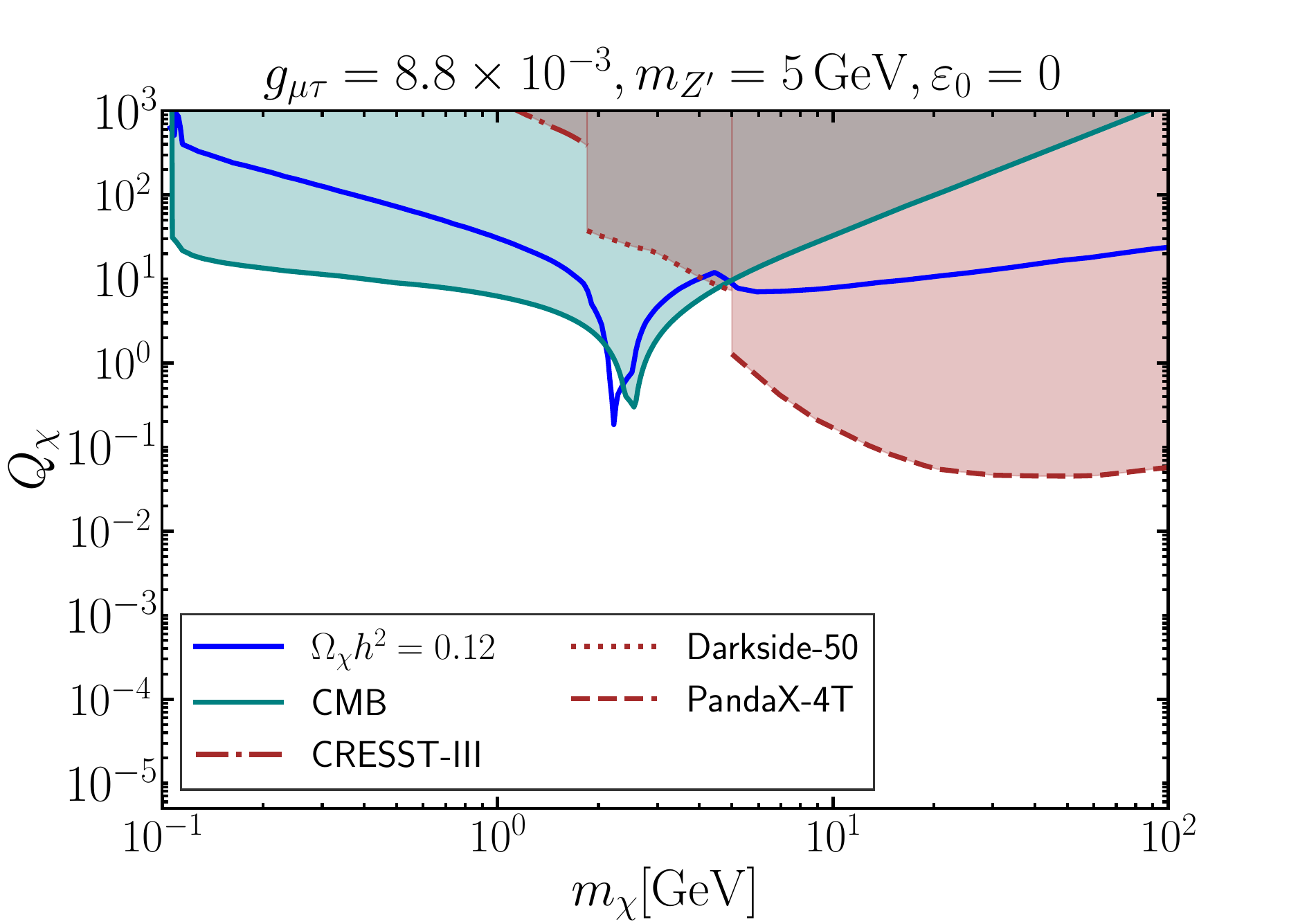}
\includegraphics[width=8.4cm,height=6.5cm]{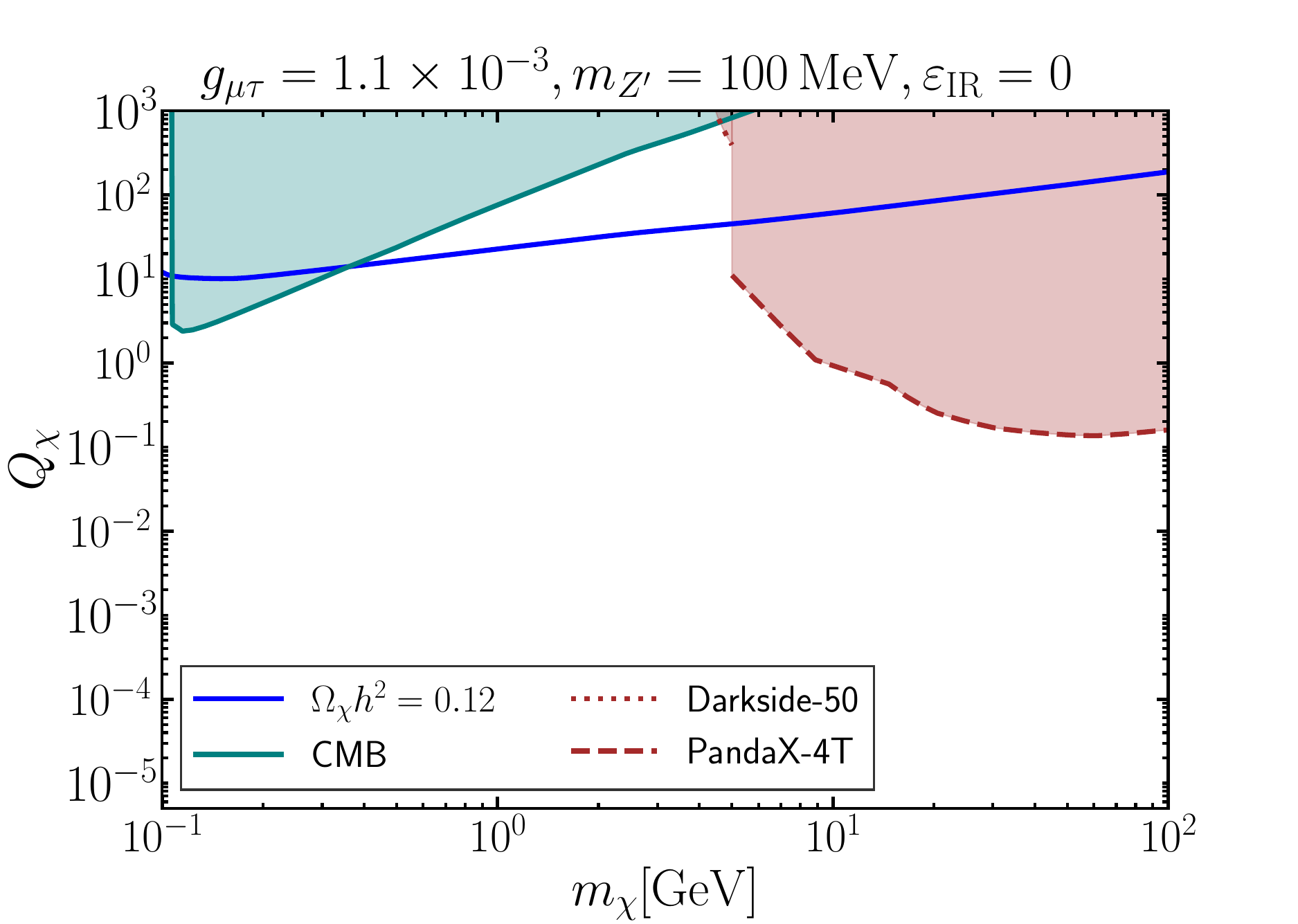}\!\!\!\!\!\!\!\!\!
\includegraphics[width=8.4cm,height=6.5cm]{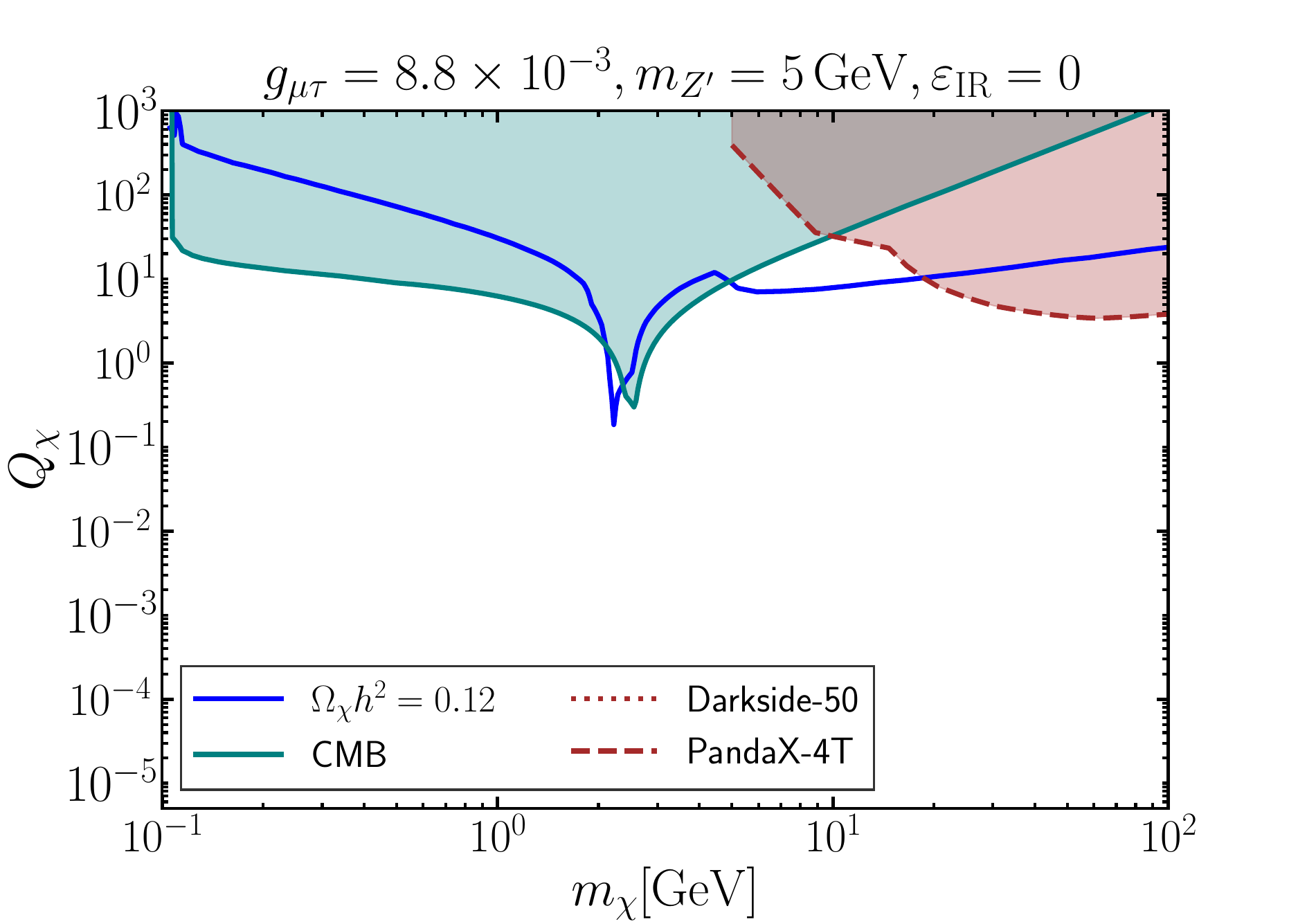}
\caption{Existing CMB constraints \cite{Planck:2018vyg,Dutta:2022wdi} (shaded dark green) and DM direct detection constraints from CRESST-III~\cite{CRESST:2019jnq}, Darkside-50~\cite{DarkSide:2018bpj} and PandaX-4T~\cite{PandaX-4T:2021bab} (shaded red) for an $L_\mu-L_\tau$ portal dark matter candidate, shown in the $Q_\chi-m_\chi$ plane, assuming $\{g_{\mu\tau} = 1.1 \times 10^{-3}, \, m_{Z^\prime} = 100\,{\rm MeV}\}$ (left panels) and $\{ g_{\mu\tau} = 8.8 \times 10^{-3}, \, m_{Z^\prime} = 5\,{\rm GeV}\}$ (right panels). Because the direct detection constraints depend on the kinetic mixing parameter, we show results for $\varepsilon_0=0$ (upper panels) and $\varepsilon_{IR}=0$ (lower panels). The relic density contours are shown as blue solid lines. }
\label{fig:NScapLmuLtaufig}
\end{figure}

\subsubsection{Relic Density}
\label{sec:DMrelic}

We assume the DM $\chi$ to be kept in thermal equilibrium with the SM plasma in the early Universe via the processes $\chi\bar{\chi} \rightarrow l^+l^-, \bar{\nu}_l\nu_l$ with $l=\mu,\tau$, and also via $\chi\bar{\chi} \rightarrow Z^\prime Z^\prime$ if $m_{Z^\prime} < m_\chi$. When the temperature of the SM plasma becomes too low to keep $\chi$ in thermal equilibrium, the $\chi$ density freezes-out, which sets the present-day DM relic density. The thermally-averaged cross-sections for relevant DM annihilation processes are 
\begin{eqnarray}
\label{eq:DMannll}
\langle \sigma v \rangle_{\chi\bar{\chi} \rightarrow l^+l^-} &=& \frac{Q^2_\chi g^4_{\mu\tau}}{\pi} \frac{m^2_\chi}{(4m^2_\chi - m^2_{Z^\prime})^2+m^2_{Z^\prime}\Gamma^2_{Z^\prime}}\left(1+\frac{m^2_l}{2m^2_\chi}\right)\sqrt{1-\frac{m^2_l}{m^2_\chi}},\\
\langle \sigma v \rangle_{\chi\bar{\chi} \rightarrow \nu_l\bar{\nu}_l} &=& \frac{Q^2_\chi g^4_{\mu\tau}}{2\pi} \frac{m^2_\chi}{(4m^2_\chi - m^2_{Z^\prime})^2+m^2_{Z^\prime}\Gamma^2_{Z^\prime}},\\
\langle \sigma v \rangle_{\chi\bar{\chi} \rightarrow Z^\prime Z^\prime} &=& \frac{Q^4_\chi g^4_{\mu\tau}}{16\pi m^2_\chi} \left(1-\frac{m^2_{Z^\prime}}{2m^2_\chi} \right)^{-2} \left(1-\frac{m^2_{Z^\prime}}{m^2_\chi}\right)^{3/2},
\label{eq:DMannZZ}
\end{eqnarray}
where the decay width of the $Z^\prime$ is given by,
\begin{eqnarray}
\Gamma_{Z^\prime} &=& g^2_{\mu\tau}\frac{m_{Z^\prime}}{12\pi}\left[ 1+ \sum_{l=\mu,\tau}\left( 1+\frac{2m^2_l}{m^2_{Z^\prime}}\right)\sqrt{1-\frac{4m^2_l}{m^2_{Z^\prime}}}\right].
\end{eqnarray}

We used \texttt{micrOmegas}$\_$v6~\cite{Belanger:2001fz,Alguero:2023zol} to compute the DM relic density contours for our model. From the expressions of the DM annihilation cross-sections in eqs.~\ref{eq:DMannll}-\ref{eq:DMannZZ} it is clear that the thermal relic density of the DM always goes as $g^{-4}_{\mu\tau}$ (when other parameters are fixed), while the dependence on the $L_\mu-L_\tau$ charge $Q_\chi$ varies depending on the dominant annihilation channel. Hence, we choose $g_{\mu\tau}$ to be fixed at two representative values (shown by red stars in fig.~\ref{fig:Zprimecons}) and investigated the variation of the DM relic density ($\Omega_\chi h^2$) in the $Q_\chi-m_\chi$ plane. Note that, while the parameter $Q_\chi$ is in principle completely free, natural values of this charge are expected to be  $Q_\chi\sim\mathcal{O}(1)$. Hence, in Fig.~\ref{fig:NScapLmuLtaufig}, we show relic density contours (solid blue lines) in the $Q_\chi-m_\chi$ plane for $\{g_{\mu\tau} = 1.1 \times 10^{-3}, \, m_{Z^\prime} = 100\,{\rm MeV}\}$ (left panels) and $\{ g_{\mu\tau} = 8.8 \times 10^{-3}, \, m_{Z^\prime} = 5\,{\rm GeV}\}$ (right panels). We find that whenever $m_\chi > m_{Z^\prime}$, the DM relic density is dictated by $\chi\bar{\chi} \rightarrow Z^\prime Z^\prime$ and hence $\Omega_\chi h^2$ goes as $Q_\chi^{-4}$ (see eqs.~\ref{eq:DMannll}-\ref{eq:DMannZZ}); otherwise, it is determined by $s$-channel mediated $\chi\bar{\chi} \rightarrow l^+l^-,\nu_l \bar{\nu}_l$ processes  and $\Omega_\chi h^2 \propto Q_\chi^{-2}$ (see eqs.~\ref{eq:DMannll}-\ref{eq:DMannZZ}). In the right panel, the dip in the relic density contour around $m_\chi \sim 2.5\,{\rm GeV}$ is due to the mediator $Z^\prime$ going on-shell. 

\subsubsection{CMB constraints}
\label{sec:DMCMB}

DM annihilation to leptonic final states (e.g., $\mu^+\mu^-$, $\tau^+\tau^-$) can alter the ionization history of of the Universe, thereby perturbing the CMB anisotropies. The Planck measurements of the CMB anisotropies~\cite{Planck:2018vyg} put strong constraints on the amount of energy injection into the SM plasma during cosmic dark ages and this in turn provides upper limits on the DM annihilation cross-sections to the SM leptons. The $95\%$ C.L. upper-limit on the DM annihilation cross-section for any specific $m_\chi$ is given by~\cite{Slatyer:2015jla},
\begin{eqnarray}
\sum_{l=\mu,\tau} \epsilon^{l^+l^-}_{\rm eff}(m_\chi) \frac{\langle \sigma v \rangle_{\chi\bar{\chi} \rightarrow l^+l^-}}{m_\chi} \lesssim 4.1 \times 10^{-28}\,{\rm cm}^{3}{\rm s}^{-1}{\rm GeV}^{-1},
\label{eq:DMCMBmaineq}
\end{eqnarray}
where the weighted efficiency factors $\epsilon^{l^+l^-}_{\rm eff}(m_\chi)$ for leptonic final states are given by,
\begin{eqnarray}
\epsilon^{l^+l^-}_{\rm eff}(m_\chi) = \frac{1}{2m_\chi}\int_0^{m_\chi} \left( 2\epsilon_{e^\pm} \frac{dN^{l^+l^-}}{dE_{e^\pm}} E_{e^\pm} dE_{e^\pm} + 2\epsilon_\gamma \frac{dN^{l^+l^-}}{dE_\gamma} E_\gamma dE_\gamma\right),
\label{eq:CMBeff}
\end{eqnarray}
with $dN^{l^+l^-}/dE_{e^\pm}$ and $dN^{l^+l^-}/dE_{\gamma}$ representing the $e^\pm$ and $\gamma$-ray spectra originating from $\chi\bar{\chi} \rightarrow l^+l^-$ which are provided in~\cite{Cirelli:2010xx}. The energy injection efficiencies for $e^\pm$ and $\gamma$, i.e., $\epsilon_{e^\pm}$ and $\epsilon_\gamma$, are available in~\cite{Slatyer:2015jla}. 

We have used the results presented in~\cite{Dutta:2022wdi} and translated them into upper-limits on $Q_{\chi}$ for any given $m_\chi$, $g_{\mu\tau}$ and $m_{Z^\prime}$. These results are shown in Fig.~\ref{fig:NScapLmuLtaufig}, where the dark-green shaded regions represent parameters that are ruled out by the measurements of the CMB anisotropies made by Planck~\cite{Planck:2018vyg}. 

\subsubsection{Direct-detection constraints}
\label{sec:DMDD}

In this model, DM interactions with SM quarks, and therefore with nucleons, are generated only through kinetic mixing (see, eq.~\ref{eq:kinmixoneloop}). In this case, the event-rate for DM-nucleon scattering is given by~\cite{Hapitas:2021ilr}
\begin{eqnarray}
R(Q_\chi,m_\chi,g_{\mu\tau},m_{Z^\prime}) &=& \frac{1}{2} N_A\,M_A\,Z^2 \left( \frac{\rho_{\odot}}{m_\chi} \right) \frac{Q^2_\chi g^2_{\mu\tau} e^2}{\pi} \times \nonumber\\ 
&& \int_{E_{\rm th}}^{E_R^{\rm max}} dE_R \epsilon_{\rm eff}(E_R)\frac{\varepsilon^2(Q)\,F^2_{\rm Helm}(Q)}{(Q^2+m^2_{Z^\prime})^2} \int_{v_{\rm min}(E_R)}^{v_{\rm esc}} d^3v \frac{f(\vec{v})}{v}, 
\label{eq:DMeventrate}
\end{eqnarray}
where $N_A$ is the number of target nucleus, $M_A$ is the mass of the target nucleus and $Z$ is the atomic number of the nucleus. The quantity $E_R \simeq Q^2/2M_A$ is the recoil energy, $\rho_\odot = 0.3\,{\rm GeV}/{\rm cm}^3$ is the local DM density, and $\epsilon_{\rm eff}(E_R)$ is a detector-specific efficiency factor. We take the Helm form factor to be~\cite{PhysRev.104.1466,Duda:2006uk},
\begin{eqnarray}
F_{\rm Helm}(Q) = \frac{3j_1(Q r_n)}{Q r_n}\exp(-Q^2 s^2/2),
\label{eq:HelmFF}
\end{eqnarray}
with $r_n = 1.14 A^{1/3}\,{\rm fm}$, $s \simeq 0.9\,{\rm fm}$, where $A$ is the mass number of the nucleus. For the DM velocity distribution we use
\begin{eqnarray}
f(\vec{v}) & \propto & \exp\left(-\frac{|\vec{v}+\vec{v}_\odot|^2}{v^2_0} \right)
\end{eqnarray}
where $v_\odot = 220\,{\rm km}/{\rm s}$, $v_0 = 235\,{\rm km}/{\rm s}$ and  $v_{\rm esc} = 550\,{\rm km}/{\rm s}$~\cite{Lin:2019uvt}. For $v_{\rm min}(E_R)$ and $E_R^{\rm max}$ we have adopted the expressions given in~\cite{Hapitas:2021ilr}. We have thus calculated the predicted event rates for various experiments by using eq.~\ref{eq:DMeventrate} together with the detector specific efficiencies provided by the CRESST-III~\cite{CRESST:2019jnq}, Darkside-50~\cite{DarkSide:2018bpj} and PandaX-4T~\cite{PandaX-4T:2021bab} collaborations.

It is important to note that the bounds derived by the experimental collaborations can be applied directly only in the limit where the mediator is heavy compared to the momentum-transfer (i.e., $m_{Z^\prime} \gg Q$) and the kinetic mixing is momentum-transfer independent, i.e., $\varepsilon(Q^2) = \varepsilon_{\rm ref}$. 
In this case the DM-proton spin-independent cross-section is well approximated by
\begin{eqnarray}
\sigma_{\chi p} \simeq \frac{Q^2_\chi g^2_{\mu\tau}}{\pi m^4_{Z^\prime}} e^2 \varepsilon_{\rm ref}^2  \mu^2_{\chi p},
\end{eqnarray}
where $\mu_{\chi p} = m_\chi m_p/(m_\chi+m_p)$ is the DM-proton reduced mass. Without loss of generality, we choose $e\,\varepsilon_{\rm ref} = \alpha g_{\mu\tau}$, and utilize the resulting upper-limits on $\sigma_{\chi p}$ (provided in~\cite{CRESST:2019jnq,DarkSide:2018bpj,PandaX-4T:2021bab}) to obtain the event-rate in the heavy-mediator limit for any given $m_\chi$: 
\begin{eqnarray}
R^{Q=0} = \frac{\sigma_{\chi p}}{2\mu^2_{\chi p}}N_A M_A Z^2 \left(\frac{\rho_\odot}{m_\chi}\right)\int_{E_{\rm th}}^{E_R^{\rm max}} dE_R \epsilon_{\rm eff}(E_R) F^2_{\rm Helm}(Q) \int_{v_{\rm min}(E_R)}^{v_{\rm esc}} d^3v \frac{f(\vec{v})}{v}.
\label{eq:DMeventrateHM}
\end{eqnarray}
This gives us the maximum number of events allowed by any given 
experiment for any particular value of $m_\chi$. 

We then compare this event-rate with the actual event-rate obtained from eq.~\ref{eq:DMeventrate}, for each chosen value of $\{ g_{\mu\tau}, \, m_{Z^\prime} \}$, to derive an upper-limit on $Q_\chi$ for any given value of $m_\chi$. The resulting limits are shown in Fig.~\ref{fig:NScapLmuLtaufig}, for $\{g_{\mu\tau} = 1.1 \times 10^{-3}, \, m_{Z^\prime} = 100\,{\rm MeV}\}$ (left panels) and $\{ g_{\mu\tau} = 8.8 \times 10^{-3}, \, m_{Z^\prime} = 5\,{\rm GeV}\}$ (right panels), for our two choices of $\varepsilon$.

From Fig.~\ref{fig:NScapLmuLtaufig}, we note that existing observations rule out a substantial part of the parameter space for which the correct DM relic density can be set via thermal freeze-out. However, there are other viable mechanisms to produce the relic density, such as the Freeze-in mechanism~\cite{Hall:2009bx}. In such cases, the relic density contours would shift downward, and hence there would be a larger region of unconstrained parameter space that is consistent with the DM relic density.

\section{Dark Matter capture in Neutron Stars}
\label{sec:ns}

We shall now determine the rate at which $L_\mu-L_\tau$ charged DM is captured in neutron stars. This will enable us to determine the parameters for which dark matter capture is maximised. Known as the geometric limit, this is the regime in which the star captures the entire incident DM flux. In this limit, the DM-induced heating of the star is maximised. This heating will have two contributions: (i) kinetic heating, which arises when the kinetic energy of the (quasi-relativistic) DM is transferred to the star through the initial scattering interaction and subsequent scattering interactions that lead to thermalisation of the DM with the NS medium and (ii) the possible annihilation of the captured DM. For reasonable choices of the NS mass and the DM density, this energy transfer will heat the NS by $\sim 2000$~K~\cite{Baryakhtar:2017dbj,Raj:2017wrv}. The observation of a neutron star below this temperature would therefore rule out parameters that fulfill the geometric capture criteria and any larger couplings.

For the capture rate calculations, we shall follow the general treatment that was developed in~\citep{Bell:2020jou,Bell:2020lmm,Anzuini:2021lnv}, the key elements of which we outline below.
The rate at which DM is captured in the star is given by
\begin{eqnarray}
C &=& \frac{4\pi}{\vstar} \frac{\rho_\chi}{m_\chi} {\rm Erf }\left(\sqrt{\frac{3}{2}}\frac{\vstar}{v_d}\right)\int_0^{\Rstar}  r^2 \frac{\sqrt{1-B(r)}}{B(r)} \sum_i \Omega_i^{-}(r)  \, dr, \label{eq:capturefinalM2text} 
\end{eqnarray}
where $v_* = 230\,{\rm km}\,{\rm s}^{-1}$ is the NS velocity, $v_d= 270\,{\rm km}\,{\rm s}^{-1}$ is the DM dispersion velocity, $\rho_\chi = 0.4\,{\rm GeV}\,{\rm cm}^{-3}$ is the DM density in the solar neighbourhood, $B(r)$ represents the time component of the Schwarzschild metric around a NS and $\Rstar$ is its radius. In eq.~\ref{eq:capturefinalM2text}, the sum extends over all possible scattering targets $i$ in the star, and $\Omega_i^-$ is the dark matter interaction rate, given by
\begin{eqnarray}
\Omega_i^{-}(r) &=& \frac{1}{32\pi^3}\int dt dE_i ds  \frac{|\overline{\mathcal{M}}(s,t,\mbeff)|^2}{s^2-[(\mbeff)^2-m_\chi^2]^2}\frac{E_i}{m_\chi}\sqrt{\frac{B(r)}{1-B(r)}}\frac{s}{\gamma(s,\mbeff)}\nn\\
&& \times\fFD(E_i,r)(1-\fFD(E_i^{'},r)),
\label{eq:intrate}
\end{eqnarray}
where
\begin{eqnarray}
\gamma(s,m_i) &=& \sqrt{(s-m_i^2-m_\chi^2)^2-4m_i^2m_\chi^2}.
\end{eqnarray}
The $\fFD$ are the distribution functions for the scattering targets, and $E_i$ and $E_i^{'}$ are the initial and final state energies, respectively, for target species $i$. The integration intervals for $s$, $t$ and $E_i$ can be found in Ref.~\cite{Bell:2020jou}. 

Note that Eq.~\ref{eq:intrate} incorporates Pauli blocking via the $(1-\fFD)$ terms. This results in an order $\mathcal{O}\left(m_\chi/p_F\right)$ suppression factor when $m_\chi\lesssim \mathcal{O}(1\GeV)$, where the fermi momentum is $p_F\sim\mathcal{O}(100\MeV)$ for leptons and $p_F\sim\mathcal{O}(400\MeV)$ for baryons. For lepton targets, $\mbeff=m_i$, where $m_i$ is the usual rest mass, while for nucleon targets the appropriate effective mass should be used \citep{Bell:2020obw,Anzuini:2021lnv} to account for strong nucleon interactions within the star. Finally, we shall consider only masses $m_\chi\gtrsim \mathcal{O}(100\MeV)$, for which collective effects of the star medium \citep{DeRocco:2022rze} can be neglected.

The squared matrix elements $|\mathcal{M}|^2$ contain effective coefficients $c_i(t)$ that depend on the momentum transfer through a momentum dependent form factor as~\citep{Bell:2020obw,Anzuini:2021lnv}\footnote{Note that, due to the running of the kinetic mixing, the factor $g_I^V$ acquires momentum dependence, differently from \citep{Bell:2020obw,Anzuini:2021lnv}.}
\begin{eqnarray}
c_i^V(t) &= g_I^V(t) F(t).
\label{eq:tdep}
\end{eqnarray}
For baryons, we have
\begin{equation}
    F(t) = \frac{1}{(1-t/Q_0^2)^4}, 
    \label{eq:formfactor}
\end{equation}
which is the square of the usual dipole form factor, and where $Q_0 \sim 1 \GeV$. For leptons, while $F(t)=1$. For purely vector-vector interactions we have (see Ref.~\citep{Anzuini:2021lnv}) 
\begin{eqnarray}
g_\mu^V(t)\sim  g_{\mu\tau}^2,    
\end{eqnarray}
while for electrons and nucleons
\begin{eqnarray}
    g_e^V(t) &=& \left(Q_e e \varepsilon(-t)\right)^2 = e^2 \varepsilon^2(-t),\\
    g_{N_i}^V(t) &=& \left(\sum_{q\in N_i}Q_q e \varepsilon(-t) \right)^2 = Q_{N_i}^2 e^2 \varepsilon^2(-t).
\end{eqnarray}
The squared matrix element is then given by
\begin{eqnarray}
|\overline{\mathcal{M}}|^2&=&\frac{Q^2_\chi g_{\mu\tau}^2 c_{i}^V}{(t-M_{Z^\prime}^2)^2} 2 \frac{2 \left(\mu ^2+1\right)^2 m_{\chi }^4-4 \left(\mu ^2+1\right) \mu ^2 s m_{\chi }^2+\mu ^4 \left(2 s^2+2 s t+t^2\right)}{\mu^4}
\end{eqnarray}
with $\mu=m_\chi/\mbeff$. 
Note that this expression can be obtained directly from the expressions in the EFT limit with the replacement
\begin{eqnarray}
    \frac{1}{\Lambda^2}\rightarrow\frac{Q_\chi\,g_{\mu\tau}^2}{t-M_{Z^\prime}^2}
\end{eqnarray}
For $M_{Z^\prime}\gg |t_{min}|\sim p_F$, and neglecting the running mixing, one recovers the EFT limit.

\begin{figure}[htb!]
\centering
\includegraphics[width=8.4cm,height=6.5cm]{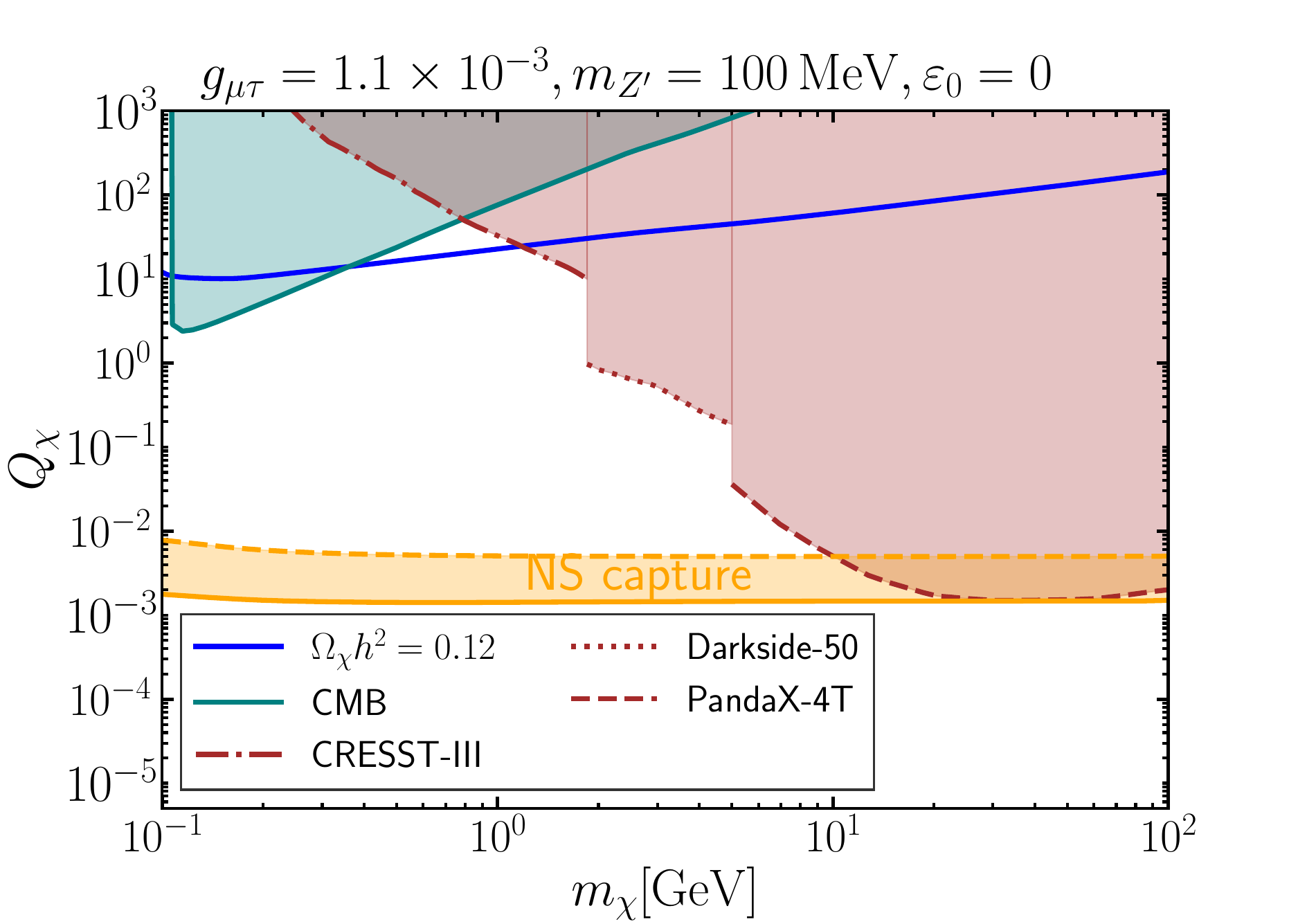}\!\!\!\!\!\!\!\!\!
\includegraphics[width=8.4cm,height=6.5cm]{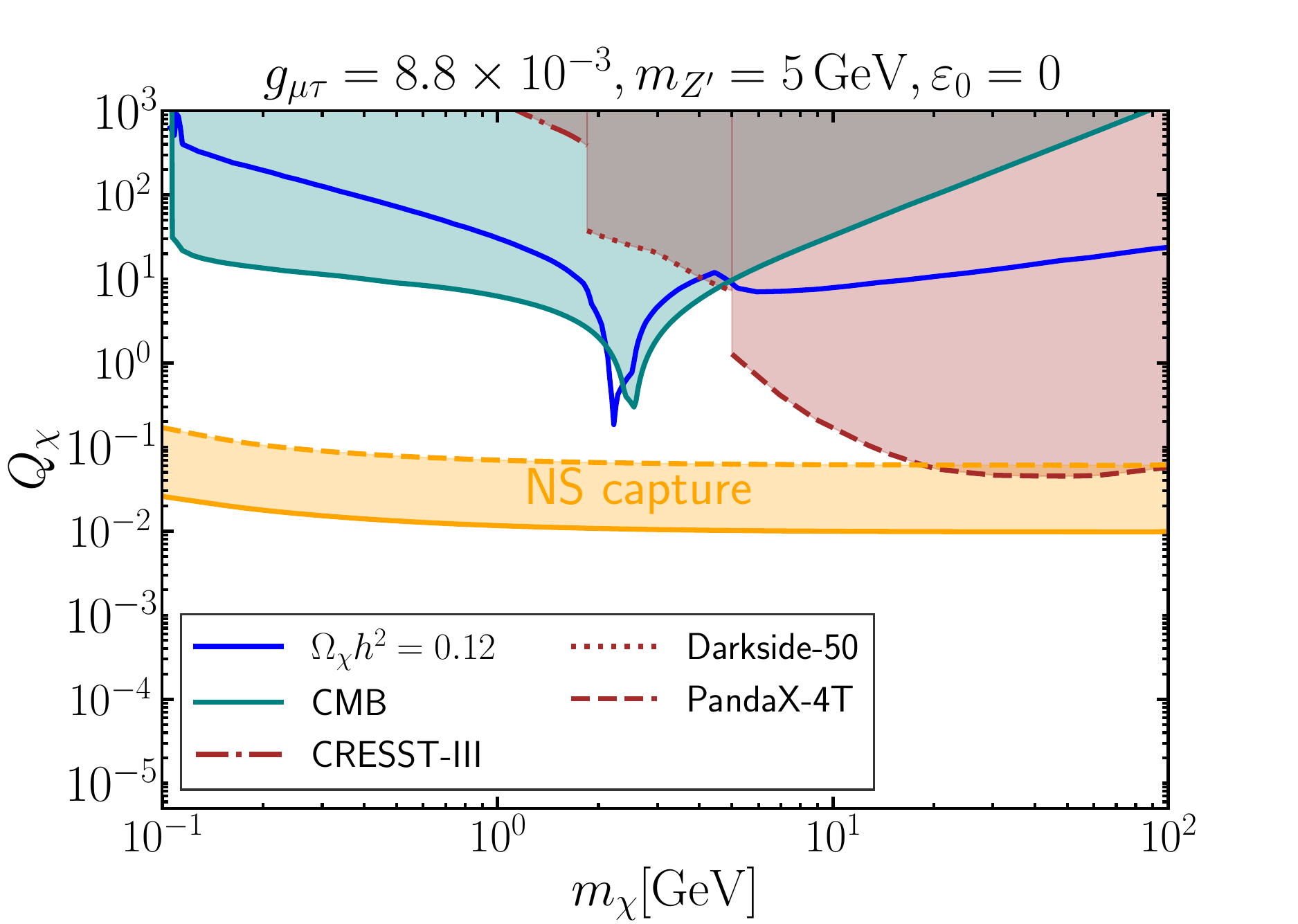}
\includegraphics[width=8.4cm,height=6.5cm]{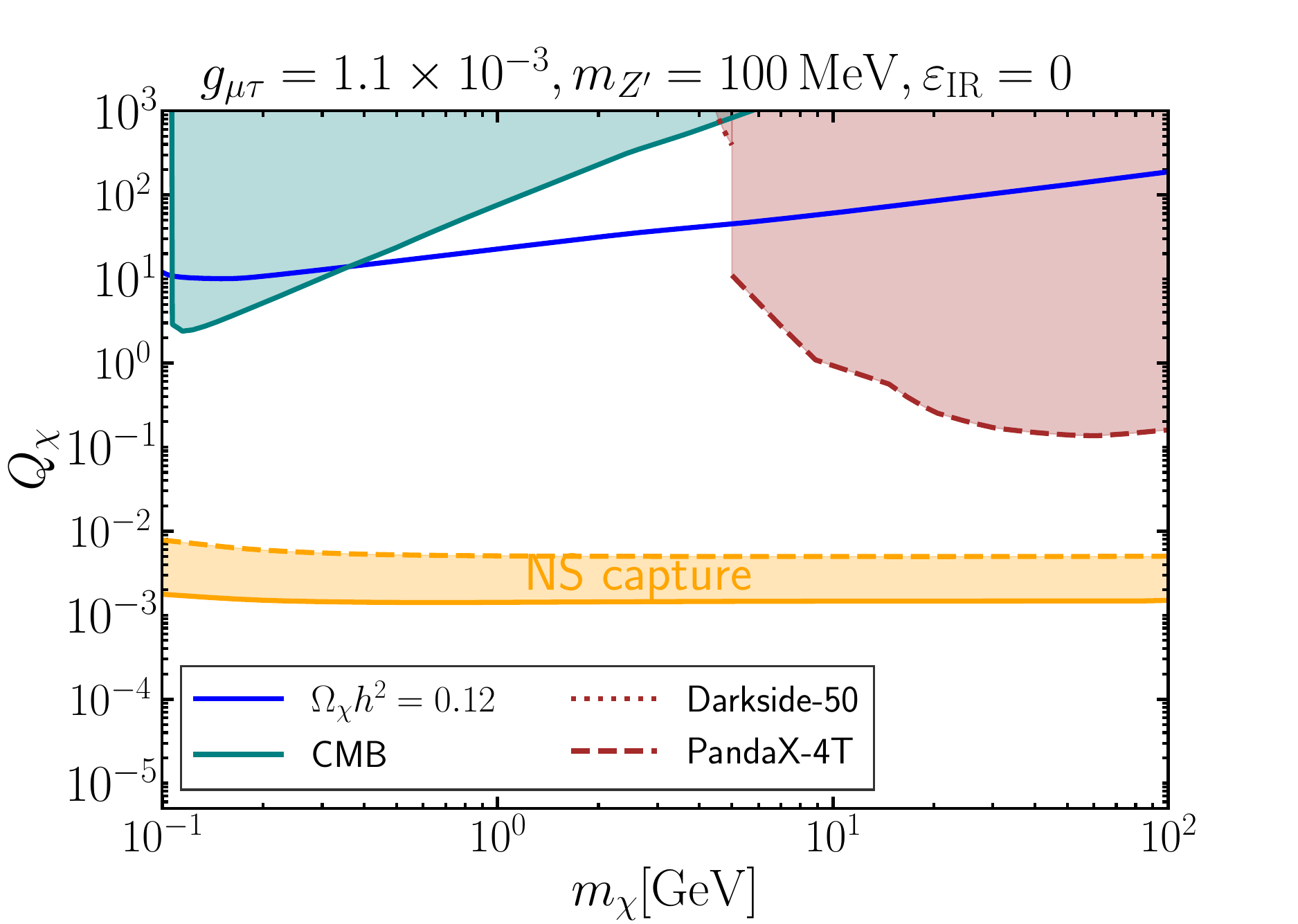}\!\!\!\!\!\!\!\!\!
\includegraphics[width=8.4cm,height=6.5cm]{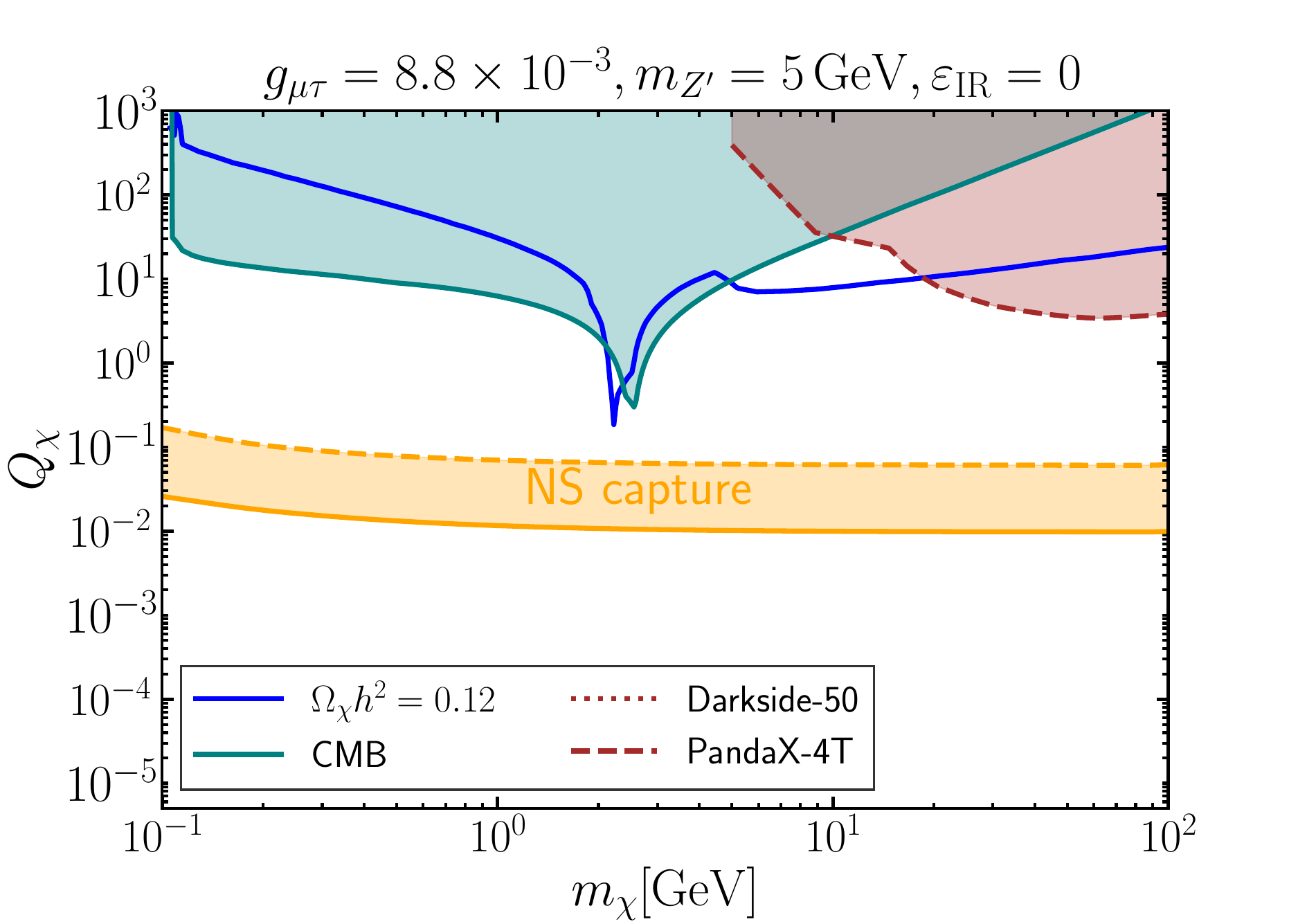}
\caption{Projected sensitivities for neutron star captures (shown by orange bands) for a $L_\mu-L_\tau$ portal dark matter candidate in the $Q_\chi-m_\chi$ plane, assuming $\{g_{\mu\tau} = 1.1 \times 10^{-3}, \, m_{Z^\prime} = 100\,{\rm MeV}\}$ (left panels) and $\{ g_{\mu\tau} = 8.8 \times 10^{-3}, \, m_{Z^\prime} = 5\,{\rm GeV}\}$ (right panels). We set $\varepsilon_0=0$ (top panels) and $\varepsilon_{IR}=0$ (lower panels).  The dashed and solid orange lines correspond to NSs of mass $1\Msun$ and $1.9\Msun$, respectively. The CMB, direct detection constraints, and relic density contours are the same as in Fig.~\ref{fig:NScapLmuLtaufig}.}
\label{fig:NScapLmuLtaufig2}
\end{figure}

We have checked that the contributions to the capture rate from protons and electrons are subdominant in the parameter space region under consideration. In fact, for values of $\varepsilon_0$ such that $\varepsilon_0=0$ or $\varepsilon_{IR}=0$, we have
\begin{eqnarray}
    |e\varepsilon(Q^2)| \le \frac{e^2 g_{\mu\tau}}{12\pi^2}\log\left(\frac{m_\tau^2}{m_\mu^2}\right) = \frac{g_{\mu\tau}\alpha}{3\pi} \log\left(\frac{m_\tau^2}{m_\mu^2}\right),
\end{eqnarray}
and therefore
\begin{eqnarray}
    \frac{g_e^V}{g_\mu^V} = \frac{g_p^V}{g_\mu^V} &\le& \frac{\alpha^2}{9\pi^2}\log^2\left(\frac{m_\tau^2}{m_\mu^2}\right) \sim 2\times 10^{-5}.
\end{eqnarray}
Due to this large suppression, the capture rate is dominated by scattering on muon targets, with the other targets contributing at less than the $1\%$ level.

As benchmark NS configurations, we consider the QMC-1 and QMC-4 models of \citep{Anzuini:2021lnv}, for which the corresponding NS mass is $1\Msun$ and $1.9\Msun$, respectively. While the NS constituents include heavy baryons in the QMC-4 case, they make a negligible contribution to the capture rate, similarly to protons. Moreover, as the total amount of DM accumulated in the NS is small compared to the number of SM particles in the star, we also neglect self-capture. Therefore, the total capture rate is determined by the $\mu$ targets alone.


Assuming black-body radiation, the DM will induce the NS to reach an equilibrium temperature of 
\begin{equation}
        T^\infty_{\chi}  = \left[\frac{B^2(\Rstar)}{4\pi\sigma_\text{SB} R_\star^2} \dot{E}_{\chi}  \right]^{1/4},
\label{eq:Tkin} 
\end{equation}
where $\sigma_\text{SB}$ is the Stefan-Boltzmann constant, $\dot E_{\chi}$ is the rate of energy deposition and is given by the sum of the kinetic and annihilation heating~\cite{Bell:2023ysh},
\begin{equation}
    \dot E_{\chi} \simeq \frac{m_\chi}{\sqrt{B(0)}} C_{\rm geom}f, \label{eq:massheating}
\end{equation} 
 where $C_{\rm geom}$ is the capture rate in the geometric limit (all incident DM captured) and we have assumed the absence of any other source of heating. The coefficient $f$ represents the fraction of the total incident DM flux captured by the star; $f=1$ in the region of parameter space above the threshold cross section, and scales linearly with the scattering cross section for smaller values.

The NS heating is maximised when the entire DM flux incident on the star is captured ($f=1)$.  In this limit, the maximum temperature obtained from kinetic (kinetic+annihilation) heating is $\sim 1510$ K ($\sim 2160$ K)  for $1\Msun$ NS and $\sim 2240$ K ($\sim 2640$ K) for $1.9\Msun$ NS. 
These temperatures do not vary appreciably for $\mathcal{O}(1)$ variations of the DM density, NS speed and mass.


In Fig. \ref{fig:NScapLmuLtaufig2} we show the potential sensitivity to NS capture. NS heating is maximised for parameters which lie on the orange lines or above. (Note that we have plotted orange bands, rather than lines, to illustrate the dependence on the NS mass.) For parameter points below the orange lines, the DM capture and hence the NS heating is sub-maximal, and thus observational prospects would be challenging. Thus, the parameter space {\it above} the orange band is the one for which this technique is of interest. Specifically, the observation of an old cold NS would enable us to exclude the entire parameter region above these bands.  Importantly, this probes the interesting region $Q_\chi\sim 1$. 
The direct detection constraints are comparable to the potential NS sensitivity only for $m_\chi \gtrsim \mathcal{O}(10\,{\rm GeV})$ with $\varepsilon_0 = 0$. In all other cases, NSs provide much greater sensitivity.

\section{Summary and Conclusion}
\label{sec:conclusion}

Kinetic heating of neutron stars via the scattering of DM particles provides a promising avenue to probe the strength of DM interactions with ordinary matter. Due to the high DM scattering rate in the dense NS medium, even for relatively small couplings, the observation of an old cold NS in a DM rich environment would rule out substantial regions of parameter space that cannot be probed by terrestrial direct detection experiments.

In this paper, we have considered a Dirac fermion DM candidate that is charged under a $U(1)_{L_\mu-L_\tau}$ gauge group and interacts with the SM via the $U(1)_{L_\mu-L_\tau}$ gauge boson, $Z^\prime$. This scenario is difficult to probe with direct detection experiments, because the $Z'$ does not couple directly to nucleons or electrons. Instead, scattering in direct detection experiments is induced only via kinetic mixing of the $Z'$ with the SM photon. As a result, the direct detection constraints are weak for a substantial region of parameter space. Therefore, the presence of a significant muon component in neutron stars provides an important avenue to test this model. 

Taking the existing constraints on $g_{\mu\tau}$ and $m_{Z^\prime}$ into account, we identified the allowed region of DM parameter space in the $Q_\chi-m_\chi$ plane that is consistent with the existing data from CMB observations~\cite{Planck:2018vyg}, direct-detection experiments~\cite{CRESST:2019jnq,DarkSide:2018bpj,PandaX-4T:2021bab} and the DM relic density. We then calculated the projected sensitivity of NS observations for this DM candidate, considering QMC-1 and QMC-4 benchmark NS configurations~\cite{Anzuini:2021lnv}. Our DM capture rate in NS calculations incorporate the effect of relativistic kinematics and Pauli blocking, which both make a substantial impact on the rates, given that the muon scattering targets in NSs are relativistic and have significant chemical potentials. Our results show that for reasonable values of $m_{Z^\prime}$ and $g_{\mu\tau}$, the observation of old cold NSs would probe substantial unexplored regions of DM parameter space for DM masses in the range $100\,{\rm MeV}-100\,{\rm GeV}$. This provides a promising way to probe these $U(1)_{L_\mu-L_\tau}$-portal DM models, for which terrestrial direct detection prospects are limited.


\section*{Acknowledgements}
This work was supported by the Australian Research Council through the ARC Centre of Excellence for Dark Matter Particle Physics, CE200100008. 

\appendix


\label{Bibliography}

\lhead{\emph{Bibliography}} 

\bibliography{Bibliography} 

\end{document}